\documentclass[amsmath,amssymb,lengthcheck,aps,prl,groupedaddress, floatfix] {revtex4}
\usepackage[utf8]{inputenc}
\usepackage{graphics}
\usepackage{graphicx}
\usepackage{latexsym}
\usepackage{epsfig}
\usepackage{amsfonts}
\usepackage[colorlinks,allcolors=blue]{hyperref}
\usepackage{afterpage}
\usepackage{array,longtable}
\usepackage{mathtools}

\begin{document}
 
\title{Study of polarization for even-denominator fractional quantum Hall states in SU(4) Graphene}

\author{Moumita Indra \footnote{Orcid ID: 0000-0002-7900-7947}}\email{moumitaindra@ee.iitb.ac.in}
\affiliation{Department of Electrical Engineering, Indian Institute of Technology (IIT) Bombay, Mumbai - 400076, India}
\author{Dwipesh Majumder \footnote{Orcid ID: 0000-0002-1221-2110}}\email{dwipesh@physics.iiests.ac.in}
\affiliation{Department of Physics, Indian Institute of Engineering Science and Technology, Shibpur, Howrah - 711103, India}

\date{\today}

\begin{abstract}
We have focused on studying the even-denominator fractional quantum Hall (EDFQH) states observed in monolayer graphene. In this article, we have studied polarization mainly for the two EDFQH states at filling fractions $\nu = 1/2, 1/4$, which are observed in an experimental study [{\color{blue} Nat. Phys. 14, 930 (2018)}] a few years ago. We have applied Chern-Simon's gauge field theory to explain the possible variational wave functions for different polarized states and calculated their ground state energies using the Coulomb potential. We have chosen the lowest energy states using suitable combinations of attached flux quanta to the electrons for different polarized states of those EDFQH states.   
\end{abstract}

\maketitle 

\section{Introduction}
Strongly correlated two-dimensional electron system under a strong perpendicular magnetic field shows a remarkable
collective phenomenon, the fractional quantum Hall effect (FQHE) \cite{Stormer, Laughlin'83}. Composite Fermion (CF) theory
\cite{CF, JainBook} beautifully explains the fractional quantum Hall (FQH) states in detail qualitatively. CF is basically
the bound state of an electron and an even number ($2p$) of quantized vortices that produce a Berry phase of $2p$ for a
closed loop around it, such that the CFs obey the same statistics as that of electrons. So the CFs experience a reduced
amount of magnetic field $B^* = B - 2p \rho \phi_0$, where $\rho$ is the electron density and $\phi_0$ is the flux quanta.
The eigenstates of the Hamiltonian of a CF are called $\Lambda$ levels, which are similar to the Landau levels (LLs) with
reduced degeneracy. $n$ number of filled $\Lambda$ levels of non-interacting CFs system (integer quantum Hall effect of CF) describes the FQHE of
filling fraction $\nu = \frac{n}{2pn \pm 1}$ of a strongly interacting electronic system. This is the beauty of this mean-field
theory. For many years, understanding the spin polarization of the various FQH states has become a topic of great interest among
experimentalists and theorists. In sufficiently high magnetic fields, the low-energy states are of course fully spin-polarized.
The partially polarized and unpolarized states occur due to the competition between the cyclotron energy and Zeeman energy.
The magnetic field (Zeeman energy) dependency of the polarization has been reported by Kukushkin and co-authors \cite{Kukushkin}
in detail. CF theory beautifully explains most of the polarized states, except for some partially polarized states such as $50\%$
polarized state of filling fraction $2/5, 2/3$ and $25\%$, $75\%$ polarized state of filling fraction $4/7, 4/9$ etc.
Ganapathy Murthy explained the partially polarized state of $2/5$ filling fraction as the density wave ordered state of
CF \cite{PRL80_1999}. Some of the partially polarized states are explained as mixture of different species of CF \cite{PLA_MI, Goerbig}. 

Besides the two-dimensional (2D) semiconductor device, FQHE has been observed in monolayer \cite{monolayer, monolayer1}, and bilayer
graphene \cite{bilayer, bilayer1}. The effective low-energy theory of the Graphene system can be described in terms of Dirac-like 
quasi-particles \cite{RMP81_109}, which endows the electron wavefunctions with an additional quantum number, termed
valley isospin, that combined with the usual electron spin, yields four-fold degenerate LLs \cite{Nat438, Zhang_2005}.
This additional symmetry adds up some new incompressible FQH states \cite{PRL96_2006, PRB74_2006, Dean_Nat7, PRB77_2008}
in graphene. Graphene has two types of Zeeman energies such as the spin Zeeman energy and valley Zeeman energy. In this SU(4) system CF theory has been used to study the polarization \cite{TCJKJ2007} and collective excitation
\cite{TCJKJ2012} by the Jain group. But the CF theory failed to explain all the possible polarized states in Graphene. Moreover, it can not explain the SU(4) polarized states of $\nu = n/(2pn \pm 1)$ with $n$ less than $3$. Goerbig and Regnault \cite{Goerbig} applied generalized Halperin's wave function to study the ground state and excited
states of $\nu = 2/3$ filling fraction in this SU(4) symmetric state.  The polarization of the FQHE in the SU(4) symmetric system has been addressed by Mandal and coworkers \cite{SM_SSM} using Chern Simon's (CS) gauge field theory \cite{PRB44_1991}.

When graphene is placed in a strong perpendicular magnetic field, a plethora of quantum Hall states are observed.
Particularly notable is the first observation of incompressible even-denominator fractional quantum Hall (EDFQH) states
at $\nu = \pm 1/2$ and $\pm 1/4$ \cite{Zibrov_Nat14}; although some EDFQH states were observed previously in higher LLs of single-component systems at $\nu = 5/2$ in GaAs \cite{PRL59_1987}, and at $\nu = 3/2 $ and $7/2$ in ZnO \cite{Falson_Nat11}. The most discussed EDFQH state in the higher LL occurs at the filling fraction $\nu = 5/2$, which can be described by the Moore Read Pfaffain \cite{Moore_Read} state.
Some EDFQH states are observed for the lowest LLs both in double quantum wells \cite{PRL68_1383} and wide single quantum wells at $\nu = 1/2$
\cite{PRL68_1379, PRL72_3405} and at $\nu = 1/4$ \cite{PRL101_2008, PRL103_2009}. The $\nu = 1/4$ state may be interpreted
as the pairing of CFs \cite{TCJKJ_2007} as described for $\nu = 1/2$ state in a thick two-dimensional electron system \cite{PRB58_1998}. In literature,
$\nu = 1/2$ state has been shown to have a large overlap with the so-called \{3,3,1\} wave function \cite{PRB47_1993, RegnaultPRB77},
which was first introduced by Halperin \cite{Halperin1983}. Beyond this \{3,3,1\} state this \{m,m,n\}  model can be applied for $\nu = 1/4$ state
considering two-component system. Possible wave function at $\nu = 1/4$ state is \{5,5,3\}, \{6,6,2\}, \{7,7,1\}. Since in monolayer graphene
there are 4-isospin configurations (two valleys and two spins) \cite{TChakrabarti, TCJKJ2006, PRB75_2007, TCJKJ2007, Ajit}, we have many
possibilities of polarized states such as spin polarization, valley polarization, and a combination of spin and valley polarization, which
we call mixed polarization. 
Golkmen \textit{et. al.} \cite{Golkmen2010} observed a marked contrast between the spin and valley degrees of freedom. They found that an electron scattering from one valley to
another requires a large momentum transfer whereas an electron scattering from one spin to another within the same valley requires a small
momentum transfer. It is also confirmed by another study by Papic \cite{Papic2010}. Wu {\it et. al.} \cite{Wu_PRB106} proposed that $\nu = 1/2$ states can be understood using the two-component parton wave-function. Recently collective excitation is studied by Sujit Narayanan \cite{Sujit_2022} for various FQH states in monolayer graphene. They have focused to study the chiral symmetry breaking orders i.e. anti-ferromagnetism and charge density orders.
We are interested here to study various possible wave functions for SU(4) graphene using CF or CS theory of multicomponent FQH system. The various possibility of incompressible EDFQH
states of filling fractions $\nu=1/2, 1/4$ in monolayer graphene was noted already in Ref. \cite{SM_SSM, PRB98_2018}. Following their work, we have investigated different polarized states of the filling fractions $\nu = 1/2, 1/4$ and then calculated the ground state energies of those states using Coulomb potential.
Although in Van der Waals heterostructures such as graphene, the gate-screening effect is very significant but we haven't included that in our calculation. We have restricted our calculation in 2D monolayer graphene, the finite width correction in this calculation will give a more accurate comparison with experiment although the basic discussions will be the same.
\section{Calculation proccedure}
Following the Chern-Simon's gauge field theory in SU(4) system, we have seen that the effective field depends on the quantum numbers
(Spin and Valley) carried by the quasiparticles. The quasiparticles of species $\alpha$ in CS theory experiences a mean effective magnetic field, $B^*_\alpha$ \cite{SM_SSM}
\begin{equation}
 B^*_\alpha = B - \phi_0 {\cal{K}}_{\alpha \beta}  \; \rho_ \beta 
\end{equation}
where $B$ is the applied actual magnetic filed, $\phi_0$ is the flux quanta and $\rho_\beta$ represents density of the electrons for the species $\beta$.
We label the valley-spin configurations $(+ \uparrow ), (+ \downarrow ), (- \uparrow ), (- \downarrow )$ by the index $\alpha = 1, 2, 3, 4$ respectively. The indices $\alpha, \beta $ runs from $1$ to $4$. We chose the symmetric $4 \times 4$ matrix ${\cal K}_{\alpha \beta}$ as follows: 
\begin{equation}
\cal{K} = \left(
      \begin{array}{llll}
         2k_1 & m_1 &n_1 & n_2 \\
         m_1 & 2k_2 & n_3 & n_4 \\
         n_1 & n_2  & 2k_3 & m_2 \\
         n_3 & n_4 & m_2 & 2k_4
      \end{array}
\right), \label{K-matrix}
\end{equation}
where $k$'s, $m$'s and $n$'s are positive integers including zero. In Ref. \cite{SM_SSM} flux attachment scheme is represented schematically.
As suggested by them we have also simplified the flux attachment that $k_1 = k_2, k_3=k_4, n_1 = n_2 = n_3 = n_4 = n$. Considering mean-field
approximation, the electron filling factor $\nu$ and the effective filling factor $\nu_\alpha$ of CS quasiparticles are related by,
\begin{equation}
\rho_\alpha / \nu_\alpha = \rho / \nu - {\cal K}_{\alpha\beta} \rho_\beta \;.
\label{eff}
\end{equation}
\subsection*{Selection of number of particles::}
We have considered that $N$ number of correlated electrons resides on the spherical surface \cite{JainBook, Fano's} in presence of radial
magnetic field created by a Dirac monopole at the center of the sphere with monopole strength $Q$ and there are $N_1$, $N_2$, $N_3$, $N_4$
number of electrons in species $1,2,3,4$ respectively. 
\begin{eqnarray*}
N&=& N_1+N_2+N_3+N_4 \\
  S&=&\frac{N_1+N_3-N_2-N_4}{N_1+N_2+N_3+N_4} \\
    V&=&\frac{N_1+N_2-N_3-N_4}{N_1+N_2+N_3+N_4} \\
      M&=&\frac{N_1+N_4-N_2-N_3}{N_1+N_2+N_3+N_4} 
\end{eqnarray*}
We have chosen the values of $N_1$, $N_2$, $N_3$, $N_4$ such a way that it will fix the values of $S$, $V$, $M$ in the following tables.
The monopole strength corresponding to the effective field as seen by species $\alpha$ can be written as,
\begin{equation}
  2q_\alpha = 2Q -  \sum_{\beta} (N_\beta-\delta_{\alpha \beta})\cal{K}_{\alpha \beta}
\end{equation}
\begin{eqnarray*}
  \sum_{\alpha} 2 q_\alpha &=& 8 Q - \sum_{\alpha} \sum_{\beta} \left ( N_\alpha -\delta_{\alpha \beta} \right ) \cal{K}_{\alpha \beta}\\
\Rightarrow Q &=& \frac{1}{8} \left (\sum_{\alpha} 2q_\alpha + \sum_{\alpha} {\cal K}_{\alpha\alpha} (N_\alpha-1) \right) \\
           &+&\frac{1}{8} \left(m_1 N_1+m_1 N_2+m_2 N_3 +m_2 N_4+ 2nN \right)
\end{eqnarray*}
2$Q\phi_0$ is the total flux through the spherical surface of radius $R$, if $B$ is the corresponding magnetic filed then
\begin{eqnarray*}
  2Q \phi_0 = 4\pi R^2 B
  \Rightarrow R^2 = \frac{2Q}{4\pi} \frac{hc}{e} \frac{e l^2}{\hbar c}  = Q {l_0}^2
\end{eqnarray*}
So the radius of the sphere is $R = \sqrt{Q}$ in unit of magnetic length $l_0$.
\subsection{Wave function}
Let us suppose that there are $\nu_1$ number of spin-up $(+)$ valley $\Lambda$-level, $\nu_2$ number of spin-down $(+)$ valley
$\Lambda$-level, $\nu_3$ number of spin-up $(-)$ valley $\Lambda$-level, $\nu_4$ number of spin-down $(-)$ valley $\Lambda$-level. 
Following the route proposed by Mandal and co-workers the spin (S), the valley (V), and the mixed (M) polarization of these
quasi-particles in a given FQH states is defined as,
 \begin{eqnarray}
 S = (\rho_1 + \rho_3 - \rho_2 - \rho_4)/ \rho,  \nonumber \\
 V = (\rho_1 + \rho_2 - \rho_3 -\rho_4 )/ \rho, \nonumber \\
 M = (\rho_1 + \rho_4 - \rho_2 - \rho_3) / \rho, 
 \end{eqnarray}
So, the above equation (\ref{eff}) relates total filling factor $\nu$ in terms of those three polarization $S, V, M$ and the attached flux numbers k's , m's and n's.
\begin{eqnarray}
\gamma_1 (2 k_1 + 1/\nu_1) + \gamma_2 m_1 + \eta_1  2n = 4/\nu   \nonumber \\
\gamma_2 (2 k_1 + 1/\nu_2) + \gamma_1 m_1 + \eta_1 2n = 4/\nu    \nonumber \\
\gamma_3 (2 k_3 + 1/\nu_3) + \gamma_4 m_2 + \eta_2 2n = 4/\nu    \nonumber \\
\gamma_4 (2 k_3 + 1/\nu_4) + \gamma_3 m_2 + \eta_2 2n = 4/\nu
\label{eqn6}
\end{eqnarray}
where, $\eta_{1(2)} = 1-(+)V$, $\delta_{1(2)} = S-(+)M$, $\gamma_{1(2)} = \eta_2 + (-) \delta_2$, and $\gamma_{3(4)} = \eta_1 + (-)\delta_1$. \\

The five indices $k_1$, $k_3$, $m_1$, $m_2$, $n$ mimic the interaction strength between different CF $\Lambda$-levels. From equation (\ref{eqn6}),
it is very clear that the polarization index $S$, $V$, $M$; four species CF $\Lambda$-level filing fractions $\nu_1, \nu_2, \nu_3, \nu_4$ and the interaction
parameters $k_1$, $k_3$, $m_1$, $m_2$, $n$ altogether fixes the total filling fraction $\nu$ of the state and the particular state is symbolized as ($k_1 k_3 m_1 m_2 n$).
\subsubsection{\textbf{Interpretation of $(k_1 k_3 m_1 m_2 n)$}}
\begin{enumerate}
  \item[(i)] $2k_1$ number of flux attachment of CF in the ($+ \uparrow$) and ($+ \downarrow$) $\Lambda$-levels.
\item[(ii)] $2k_3$ number of flux attachment of CF in the ($- \uparrow$) and ($- \downarrow$) $\Lambda$-levels.
\item[(iii)] $m_1$ flux attachment in the ($+ \uparrow$) level seen by the CFs in ($+ \downarrow$) $\Lambda$-level and vice versa.
\item[(iv)] $m_2$ flux attachment in the ($- \uparrow$) level seen by the CFs in ($- \downarrow$) $\Lambda$-level and vice versa.
\item[(v)] $n$ flux attachment in the ($+ \uparrow \& \downarrow $) level seen by the CFs in ($- \uparrow \& \downarrow$) $\Lambda$-levels and vice versa.\\
\end{enumerate}

The variational wave function for the state is proposed by Mandal and coworker \cite{SM_SSM} and is given by
\begin{eqnarray}
\Psi _{\{\kappa_{\alpha \beta }\}}  &=&  {\cal P}_L \Phi_{\nu_1}(\Omega_1^1,\cdots,\Omega_{N_1}^1) \Phi_{\nu_2}(\Omega_1^2,\cdots,\Omega_{N_2}^2)  \nonumber \\
&&\times \Phi_{\nu_3}(\Omega_1^3,\cdots,\Omega_{N_3}^3) \Phi_{\nu_4}(\Omega_1^4,\cdots,\Omega_{N_4}^4) 
\nonumber \\
&&\times J_{11} \;J_{22}\; J_{33}\; J_{44}\; J_{12} \;J_{34}\; J_{13}\; J_{14}\; J_{23}\; J_{24} \nonumber 
\label{Wavefn}
\end{eqnarray}
where, $\Phi_{\nu_1}$ is the Slater determinant of $\nu_1$ filled $\Lambda$ level CFs, $\Omega_1, \cdots \Omega_{N_{\alpha}}$ are the positions of CFs on the spherical surface, upper index indicate the different species of CFs and the Jastrow factor is given by
\begin{eqnarray}
  J_{\alpha \beta} &=&  \prod_{i, j}^{N_\alpha, N_\beta} (u_i^{(\alpha)} v_j^{(\beta)}-u_j^{(\beta)} v_i^{(\alpha)})^{\kappa_{\alpha \beta }} \mbox{~~~~ if $\alpha \; \neq \beta$ }\nonumber \\
  J_{\alpha \alpha}& =&  \prod_{i< j}^{N_\alpha} (u_i^{(\alpha)} v_j^{(\alpha)}-u_j^{(\alpha)} v_i^{(\alpha)})^{\kappa_{\alpha \alpha }} \nonumber 
  \end{eqnarray}
The prefix within bracket represent the CFs of different $\Lambda$ levels. Here we have used the spinor coordinates 
\begin{eqnarray*}
  u(\Omega) = cos(\theta/2) e^{-i\phi/2} \mbox{~~~ and ~~~~} v(\Omega) = sin (\theta/2) e^{i\phi/2}
\end{eqnarray*}
Some of the above wave functions are identical with the wave function proposed by Regnault and others using the plasma picture
of the FQHE with internal SU(4)  symmetry \cite{PRB77_2008}. The ground state energy ($E_g$) per particle corresponding to a
particular state with flux attachment ($k_1 k_3 m_1 m_2 n$) can be calculated by the Monte Carlo method as
\begin{eqnarray}
  E_g &=& \frac{E + E_{bg}}{N} \nonumber \\
  &=& \left [ \frac{<\Psi _{\{\kappa_{\alpha \beta}\}} | H | \Psi _{\{\kappa_{\alpha \beta}\}}> }{<\Psi _{\{\kappa_{\alpha \beta}\}}| \Psi _{\{\kappa_{\alpha \beta}\}}>} - \frac{N^2 e^2}{2 \epsilon R} \right] /N \;
\end{eqnarray}
Here, $H = \sum_{i<j} \frac{e^2}{r_{i,j}}$ is the Coulomb interaction, with $r_{i,j}$ as the inter-electronic distance and  
$E_{bg}=\frac{-N^2 e^2}{2 \epsilon R}$ term represents the background energy i.e. the interaction between the electrons and
the background positive ions where $\epsilon$ is the dielectric constant of the background material. LL mixing parameter is
independent of the applied magnetic field in the graphene system \cite{LLmixGraphene}, so we do not need to include the
LL mixing \cite{LLmix97, LLmix2009} in our calculation.
\begin{table}[h]
\caption{Possible polarized states fitted with equation \eqref{eqn6} for filling fraction $\nu = 1/2$ and their ground state energies with different parameters}
   \label{tab1}
   \begin{center}
  \begin{tabular}[c]{|| c c c  c c |c c c| c c c c |c|| }
    \hline
   $k_1$ & $k_3$ & $m_1$ & $m_2$ & $n$ & $V$ & $S$   & $M$ & $\nu_1$ & $\nu_2$ & $\nu_3$ & $\nu_4$ & $E_g (e^2/\epsilon l_0)$ \\
   \hline
   1 & 1 & 1 & 1 & 2  &  &  &  &  &  &  &  &  -0.4677  \\ 
   1 & 1 & 3 & 3 & 1  & 0 & 0 & 0 & 1 & 1 & 1 & 1 &  -0.46315 \\
   2 & 1 & 1 & 3 & 1  &  &  &  &  &  &  &  &  -0.45576\\
   2 & 2 & 1 & 1 & 1  &  &  &  &  &  &  &  &  -0.44983 \\
   \hline
   
    \hline
     1 & 1 & 1 & 1 & 2  &  &  &  & 1 & 1 & 1 & 1 & -0.46742\\
     1 & 2 & 2 & 3 & 1  &  &  &  &  &  &  &  &  -0.46351\\
     1 & 3 & 2 & 1 & 1  & 1/3 & 0 & 0 & 1 & 1 & 1 & 1 &  -0.46012\\
     1 & 1 & 2 & 5 & 1  &  &  &  &  &  &  &  &  -0.45653\\
     \hline
     
    \hline
     &  &  &  &   &  &  &  & 1 & 1 & 0 & 0 & -0.46321 \\ 
      
     &  &  &  &   & 1 & 0 & 0 & 1 & 1 & 1 & 0 & -0.46303 \\
      
     &  &  &  &   &  &  &  & 1 & 1 & 0 & 1 & -0.46298 \\  
      
    1 & 1 & 1 & 1 & 2 & -1 & 0 & 0 & 0 & 0 & 1 & 1 & -0.46223 \\

     &  &  &  &  & 1/4 & 0 & 0 & 1 & 1 & 1 & 1 & -0.46759\\ 
     &  &  &  &  & 1/2 & 0 & 0 & 1 & 1 & 1 & 1 & -0.46671\\ 
     &  &  &  &  & 2/3 & 0 & 0 & 1 & 1 & 1 & 1 & -0.46567\\ 
     \hline
  
     \hline
     1 & 1 & 1 & 1 & 3  & 1 & 0 & 0 & 1 & 1 & 0 & 0 & -0.46321 \\
          
      &  &  &  &   & -1 & 0 & 0 & 0 & 0 & 1 & 1 & -0.46223 \\
     
     \hline
     1 & 2 & 1 & 1 & 1  & 1 & 0 & 0 & 1 & 1 & 0 & 0 & -0.53459 \\
          
     2 & 1 & 1 & 1 & 1  & -1 & 0 & 0 & 0 & 0 & 1 & 1 & -0.53353 \\
     
    \hline
  
     \hline
     1 & 2 & 1 & 1 & 2  & 1 & 0 & 0 & 1 & 1 & 0 & 0 & -0.46302 \\
          
     2 & 1 & 1 & 1 & 2  & -1 & 0 & 0 & 0 & 0 & 1 & 1 & -0.46186 \\
     
     \hline
  
     \hline 
      &  &  &  &  & 0 & 1 & 0 & 1 & 0 & 1 & 0 & -0.51384 \\
     
      &  &  &  &   & 0 & -1 & 0 & 0 & 1 & 0 & 1 & -0.51289 \\
       
     1 & 1 & 1 & 2 & 1  & 0 & 0 & 1 & 1 & 0 & 0 & 1 & -0.51381 \\
          
      &  &  &  &   & 0 & 0 & -1 & 0 & 1 & 1 & 0 & -0.51282 \\

    \hline
  
     \hline
      &  &  &  &  & 0 & 1 & 0 & 1 & 0 & 1 & 0 & -0.49495 \\
     
      &  &  &  &   & 0 & -1 & 0 & 0 & 1 & 0 & 1 & -0.49391 \\
       
     1 & 1 & 2 & 2 & 1  & 0 & 0 & 1 & 1 & 0 & 0 & 1 & -0.49514 \\
          
      &  &  &  &   & 0 & 0 & -1 & 0 & 1 & 1 & 0 & -0.49398 \\
  \hline
  
  \hline
      &  &  &  &  & 0 & 1 & 0 & 1 & 0 & 1 & 0 & -0.49501 \\
     
      &  &  &  &   & 0 & -1 & 0 & 0 & 1 & 0 & 1 & -0.49400 \\
       
     1 & 1 & 1 & 3 & 1  & 0 & 0 & 1 & 1 & 0 & 0 & 1 & -0.49492 \\
          
      &  &  &  &   & 0 & 0 & -1 & 0 & 1 & 1 & 0 & -0.49407 \\
      \hline
      
\end{tabular}
\end{center}
\end{table}
\begin{figure}
 \centering
    {\includegraphics[width=0.5\textwidth]{nu12.eps}}
   \caption{Ground state energy per particle for different combinations of different interaction parameters ($k_1 k_3 m_1 m_2 n$) with different values of polarization indices $V, S, M$ at filling fraction $\nu = 1/2$ as a function of inverse of particle number. [$N \rightarrow \infty$ (or, $1/N \rightarrow 0$) denotes the thermodynamic limt, which has been listed in Table \ref{tab1}, \ref{tab3}]}
   \label{Fig2}
   
   {\includegraphics[width=0.5\textwidth]{nu14.eps}}
   \caption{Ground state energy per particle for different combinations of different interaction parameters ($k_1 k_3 m_1 m_2 n $) with different values of polarization indices $V, S, M$ at filling fraction $\nu = 1/4$.}
   \label{Fig3}
 \end{figure}
%
\begin{table}[h]
   \caption{Lowest energy states with different polarization indices for $\nu = 1/2$}
   \label{tab2}
   \begin{center}
  \begin{tabular}[c]{|| c c c  c c |c c c| c c c c || }
    \hline
   $k_1$ & $k_3$ & $m_1$ & $m_2$ & $n$ & $V$ & $S$   & $M$ & $\nu_1$ & $\nu_2$ & $\nu_3$ & $\nu_4$  \\
    \hline
     1 & 1 & 1 & 1 & 2 & 0 & 0 & 0 & 1 & 1 & 1 & 1 \\
      &  &  &  &  & 1/3 & 0 & 0 & 1 & 1 & 1 & 1 \\
      \hline
     1 & 2 & 1 & 1 & 1 & 1 & 0 & 0 & 1 & 1 & 0 & 0  \\ 
     
     2 & 1 & 1 & 1 & 1 & -1 & 0 & 0 & 0 & 0 & 1 & 1  \\
       \hline
      &  &  &  &  & 0 & 1 & 0 & 1 & 0 & 1 & 0  \\  
      
     1 & 1 & 1 & 2 & 1 & 0 & -1 & 0 & 0 & 1 & 0 & 1  \\ 
  
      &  &  &  &  & 0 & 0 & 1 & 1 & 0 & 0 & 1  \\  
   
      &  & &  &  & 0 & 0 & -1 & 0 & 1 & 1 & 0  \\ 
          
  \hline
\end{tabular}
\end{center}
\end{table}
\begin{table}
   \caption{Possible polarized states fitted with equation \eqref{eqn6} of filling fraction $\nu = 1/4$ and their ground state energies with different parameters}
   \label{tab3}
   \begin{center}
  \begin{tabular}[c]{|| c c c  c c |c c c| c c c c |c|| }
 \hline
  $k_1$ & $k_3$ & $m_1$ & $m_2$ & $n$ & $V$ & $S$   & $M$ & $\nu_1$ & $\nu_2$ & $\nu_3$ & $\nu_4$ & $E_g (e^2/\epsilon l_0)$ \\
    \hline
     1 & 1 & 3 & 3 & 5 &  &  &  & 1 & 1 & 1 & 1 & -0.21142 \\ 
      
     2 & 2 & 1 & 1 & 5 &  &  &  & 1 & 1 & 1 & 1 & -0.17752 \\
     
     2 & 2 & 3 & 3 & 4 & 0 & 0 & 0 & 1 & 1 & 1 & 1 & -0.35856 \\
     
     2 & 2 & 5 & 5 & 3 &  &  &  & 1 & 1 & 1 & 1 & -0.35749 \\
     
     3 & 3 & 1 & 1 & 4 &  &  &  & 1 & 1 & 1 & 1 & -0.33227 \\

    \hline
     2 & 1 & 3 & 1 & 1 & 1 & 0 & 0 & 1 & 1 & 0 & 0 & -0.45215 \\ 
      
     2 & 2 & 3 & 1 & 1 & 1 & 0 & 0 & 1 & 1 & 0 & 0 & -0.45217 \\
     
    \hline
     1 & 2 & 1 & 3 & 1 & -1 & 0 & 0 & 0 & 0 & 1 & 1 & -0.45178 \\ 
      
     2 & 2 & 1 & 3 & 1 & -1 & 0 & 0 & 0 & 0 & 1 & 1 & -0.45183 \\
     
    \hline
     2 & 2 & 3 & 3 & 1 & 1 & 0 & 0 & 1 & 1 & 0 & 0 & -0.45221 \\ 
      
      &  &  &  &  &  -1 & 0 & 0 & 0 & 0 & 1 & 1 & -0.45196 \\
    \hline
      
     2 & 2 & 3 & 3 & 2 & 1 & 0 & 0 & 1 & 1 & 0 & 0 & -0.41279 \\  
      
     &  &  &  &  & -1 & 0 & 0 & 0 & 0 & 1 & 1 & -0.41257 \\
     \hline
      
     3 & 1 & 1 & 1 & 1 & 1 & 0 & 0 & 1 & 1 & 0 & 0 & -0.42212 \\  
      
     2 & 3 & 1 & 1 & 1 & -1 & 0 & 0 & 0 & 0 & 1 & 1 & -0.42164 \\
     
    \hline
       &  &  &  &  & 0 & 1 & 0 & 1 & 0 & 1 & 0 & -0.41283 \\
     
     2 & 2 & 1 & 1 & 3 & 0 & -1 & 0 & 0 & 1 & 0 & 1 & -0.4125 \\
     
       &  &  &  &  & 0 & 0 & 1 & 1 & 0 & 0 & 1 & -0.41277 \\
     
       &  &  &  &  & 0 & 0 & -1 & 0 & 1 & 1 & 0 & -0.41253 \\
     
     \hline
       &  &  &  &  & 0 & 1 & 0 & 1 & 0 & 1 & 0 & -0.40453 \\
     
     2 & 2 & 1 & 2 & 3 & 0 & -1 & 0 & 0 & 1 & 0 & 1 & -0.40423\\
     
       &  &  &  &  & 0 & 0 & 1 & 1 & 0 & 0 & 1 &  -0.40439\\
     
       &  &  &  &  & 0 & 0 & -1 & 0 & 1 & 1 & 0 & -0.40407 \\

       \hline
         
     \hline
       &  &  &  &  & 0 & 1 & 0 & 1 & 0 & 1 & 0 & -0.40445 \\
     
     2 & 2 & 2 & 1 & 3 & 0 & -1 & 0 & 0 & 1 & 0 & 1 & -0.40423 \\
     
       &  &  &  &  & 0 & 0 & 1 & 1 & 0 & 0 & 1 & -0.40444 \\
     
       &  &  &  &  & 0 & 0 & -1 & 0 & 1 & 1 & 0 & -0.40433 \\

       \hline
       &  &  &  &  & 0 & 1 & 0 & 1 & 0 & 1 & 0 & -0.39658 \\
          
     2 & 2 & 2 & 2 & 3 & 0 & -1 & 0 & 0 & 1 & 0 & 1 & -0.39634 \\
      
       &  &  &  &  & 0 & 0 & 1 & 1 & 0 & 0 & 1 & -0.39654 \\
     
       &  &  &  &  & 0 & 0 & -1 & 0 & 1 & 1 & 0 & -0.39646 \\
       
    \hline
       
\end{tabular}
\end{center}
\end{table}
%
\begin{table}[h]
   \caption{Lowest energy states of different polarization indices for $\nu = 1/4$}
   \label{tab4}
   \begin{center}
  \begin{tabular}[c]{|| c c c  c c |c c c| c c c c || }
    \hline
   $k_1$ & $k_3$ & $m_1$ & $m_2$ & $n$ & $V$ & $S$   & $M$ & $\nu_1$ & $\nu_2$ & $\nu_3$ & $\nu_4$ \\
    \hline
    2 & 2 & 3 & 3 & 4 & 0 & 0 & 0 & 1 & 1 & 1 & 1 \\ 
     
    2 & 2 & 5 & 5 & 3 &  &  &  &  &  &  &  \\  
  \hline
    2 & 1 & 3 & 1 & 1 &  &  &  &  &  &  &  \\ 
      
    2 & 2 & 3 & 1 & 1 & 1 & 0 & 0 & 1 & 1 & 0 & 0 \\
    
    2 & 2 & 3 & 3 & 1 &  &  &  &  &  &  &  \\ 
    \hline
  
    1 & 2 & 1 & 3 & 1 &  &  &  &  &  &  &  \\ 
     
    2 & 2 & 1 & 3 & 1 & -1 & 0 & 0 & 0 & 0 & 1 & 1  \\
    
    2 & 2 & 3 & 3 & 1 &  &  &  &  &  &  &  \\ 
    
       \hline
      &  &  &  &  & 0 & 1 & 0 & 1 & 0 & 1 & 0  \\  
      
     2 & 2 & 1 & 1 & 3 & 0 & -1 & 0 & 0 & 1 & 0 & 1 \\ 
  
      &  &  &  &  & 0 & 0 & 1 & 1 & 0 & 0 & 1  \\  
   
      &  & &  &  & 0 & 0 & -1 & 0 & 1 & 1 & 0 \\ 
      \hline
     
\end{tabular}
\end{center}
\end{table}
\section{Results \& Conclusion}
We observed that the EDFQH states do not correspond to the Fermi sea of CFs. 
According to the Chern-Simon's theory, we have different degeneracy for four $\Lambda$-levels. Also in our earlier study of SU(2) system \cite{PLA_MI}, we didn’t get the CF Fermi-sea using Chern-Simon’s theory.
Zibrov \textit{et. al.} \cite{Zibrov_Nat14} proposed that the EDFQH states are associated with a phase transition from a partial sublattice polarized \cite{PSP} to a canted antiferromagnet phase \cite{CAF}. With this idea, Sujit \textit{et. al.} \cite{PRB98_2018} concluded that there is a phase transition from a state with $V \neq 0, S = 0, M = 0$ to one with $V = 0$ and either $S$ or $M \neq 0$.  Sujit and their group claimed that if the flux attachment set for $\nu = 1/2$ is denoted by an index $j=k_1, k_3, m_1, m_2, n$, then for $\nu = 1/4$ the flux attachment set is found to have $j+1$ i.e. each of the indices are increased by one. This observation also agrees with our result.
In this article, we have accumulated all possible wave-functions for the recently observed EDFQH states at $\nu = 1/2, 1/4$ and calculated their ground state energies which help us to find out the befitted wave functions for those states. 

We have studied here the possible polarizations for the two EDFQH states ($\nu = 1/2, 1/4$). We have considered different
combinations of flux attachment $(k_1 k_3 m_1 m_2 n)$ between the four species of electrons and calculated their ground
state energies using the Coulomb potential for a finite number of particles. To get the thermodynamic limit, we have extrapolated
the result in the $1/N \rightarrow 0$ region. The energy values for the filling fraction $\nu = 1/2$ and $1/4$ are listed in
TABLE \ref{tab1} and TABLE \ref{tab3}. 
Energy has been expressed in natural unit $e^2/\epsilon l_0$. The error of the Monte-Carlo integration is very less.\\
\textbf{Observations on filling fraction $\nu = 1/2$:}\\
We have observed that different interaction strength between the four species can lead to different polarization. We can
have the unpolarized state of $\nu = 1/2$ for different combinations of flux attachment i.e. different interaction parameters;
keeping the four $\Lambda$-level filling fractions unity ($\nu_1=\nu_2=\nu_3=\nu_4=1$). The ground state energies differ in each
of the cases. 
\begin{enumerate}
\item[(i)] 
We found that the state with interaction $(1 1 1 1 2)$ has the lowest energy (see FIG. \ref{Fig2} (a)). So this
is the most stable state for the unpolarized condition. Other states having interaction parameters $(1 1 3 3 1)$, $(2 1 1 3 1)$,
$(2 2 1 1 1)$ have higher energy compared to that. These states can be thought of as next stable states. By a slight change in the
interaction parameters the system can suffer transition from one state to another.   
\item[(ii)] 
After that, we have considered different valley polarized states. We have observed that also for $V=1/3$ polarized state,
with other two polarization index zero ($S$, $M=0$), the state with interaction $(1 1 1 1 2)$ has the lowest energy (Fig. \ref{Fig2} (b)).
\item[(iii)] 
Then we have checked the energies for different valley polarizations for the combination $(1 1 1 1 2)$ keeping
the spin and mixed polarization zero ($S=M=0$). Energies of those valley polarized states ($V=0,1/4,1/3,1/2,2/3,\pm 1$) are
slightly differ from each other. So there is a high chance for the phase transition depending upon the Zeeman energies between
the four $\Lambda$-levels. We have noted that $V = 0$ state has the lowest energy and $V=\pm 1$ state has the highest energy
and intermediate states lie in between. 
\item[(iv)] 
Besides, we have checked whether there is any other state having lower energy for $V=\pm 1$ polarization with other two polarization
index ($S, M$) zero. We found that the states $(1 2 1 1 1)$ and $(2 1 1 1 1)$ have the lowest energy for the valley polarization
$V = 1$ and $V = -1$ respectively; with other two polarization index $S$, $M$ zero for both the two cases (see FIG. \ref{Fig2} (c)). 
\item[(v)] 
Next, we calculated the same taking the polarization $S$ (or $M$) as $\pm 1$  and the other two polarization indices $V$, $M$ (or $V$, $S$) as zero,
as suggested by Sujit {\it{et. al.}} \cite{PRB98_2018}. We observed that for the interaction $(1 1 1 2 1)$ we get the lowest energy when the polarization
$S$ (or, $M$) is $\pm 1$ and the energy values are almost equal (see FIG. \ref{Fig2} (d)). All the results for the filling fraction $\nu=1/2$ are shown in TABLE \ref{tab2}. \\
\end{enumerate}
\textbf{Observations on filling fraction $\nu = 1/4$:}\\
Similarly, we have checked all the above possibilities for the filling fraction $\nu=1/4$, the results are shown in Fig. \ref{Fig3}
and in TABLE \ref{tab3}. 
\begin{enumerate}
\item[(i)] 
For the unpolarized state ($V=S=M=0$), we have found that $(2 2 3 3 4)$, $(2 2 5 5 3)$ states have almost the same energy which is the
lowest for this polarization (see FIG. \ref{Fig3} (a)). So the unpolarized state is doubly degenerate for $\nu=1/4$. 
\item[(ii)]
We have found that for $V = 1$ polarized state with other two polarization zero ($S$, $M = 0$); $(2 1 3 1 1)$, $(2 2 3 1 1)$ and $(2 2 3 3 1)$
states have the same energy value (see FIG. \ref{Fig3} (c)). That is those three states are degenerate metastable states; so the system can
change any one of those. 
\item[(iii)]
For the valley polarization $V = -1$ with $S$, $M = 0$; there are also three degenerate metastable states, which
are $(1 2 1 3 1)$, $(2 2 1 3 1)$, and $(2 2 3 3 1)$ (see FIG. \ref{Fig3} (d)). 
\item[(iv)]
We have found that $(2 2 1 1 3)$ has the lowest energy when spin or mixed polarization $S$ (or, $M$) is $\pm 1$ and valley polarization is zero
(see Fig. \ref{Fig3} (b)). We also noticed that for same interaction parameters ($2 2 1 1 3$), change in Zeeman energies between four $\Lambda$-levels
can lead to different polarization $S=\pm 1$, $M=\pm 1$ states, those states have same energies i.e. those are the degenerate states for the filling
fraction $\nu=1/4$. The lowest energy states for the different polarized states of filling fraction $\nu = 1/4$ are enlisted in TABLE \ref{tab4}.
\end{enumerate}
In conclusion, we want to mention that we have confirmed the most stable ground state of a particular polarized state at filling fraction $\nu=1/2$ and $\nu=1/4$, and possible phase transition for different polarized states within the filling fraction, which are associated with the prediction of Sujit {\it et. al.} \cite{PRB98_2018}.

\section{acknowledgement} Moumita is thankful to the Department of Physics, IIEST Shibpur for providing her with the computing facility. \\

\end{document}